\documentclass[letterpaper,english,aip,amsmath,amssymb,reprint,jcp]{revtex4-1}
\usepackage[T1]{fontenc}
\usepackage[utf8]{inputenc}
\usepackage{babel}
\usepackage{units}
\usepackage{amsmath}
\usepackage{graphicx}
\usepackage{subscript}
\usepackage[unicode=true]
 {hyperref}

\makeatletter

\pdfpageheight\paperheight
\pdfpagewidth\paperwidth

\providecommand{\tabularnewline}{\\}

\usepackage{dcolumn}
\usepackage{bm}
\usepackage{etoolbox}

\makeatletter
\def\@email#1#2{%
 \endgroup
 \patchcmd{\titleblock@produce}
  {\frontmatter@RRAPformat}
  {\frontmatter@RRAPformat{\produce@RRAP{*#1\href{mailto:#2}{#2}}}\frontmatter@RRAPformat}
  {}{}
}%
\makeatother

\makeatother

\begin{document}
\author{Jonathon G. Gray}
\affiliation{Dept. of Chemistry and Chemical Biology, Rutgers University, Piscataway,
NJ 08854}
\author{George M. Giambaşu}
\affiliation{Institute for Quantitative Biomedicine, Rutgers University, Piscataway,
NJ 08854}
\altaffiliation{Computational and Structural Chemistry Dept., Merck Research Laboratories, 33 Avenue Louis Pasteur, Boston, MA 02115}

\author{David A. Case{*}}
\affiliation{Dept. of Chemistry and Chemical Biology, Rutgers University, Piscataway,
NJ 08854}
\author{Tyler Luchko{*}}
\affiliation{Dept. of Physics, California State University, Northridge CA, 91330}
\date{\today}

\title{Integral equation models for solvent in macromolecular crystals}

\begin{abstract}
Solvent can occupy up to $\sim70\%$ of macromolecular crystals
and hence having models that predict solvent distributions in periodic
systems could improve in the interpretation of crystallographic data.
Yet there are few implicit solvent models applicable to periodic solutes
while crystallographic structures are commonly solved assuming a flat
solvent model. Here we present a newly-developed periodic version
of the 3D-RISM integral equation method that is able to solve for
efficiently and describe accurately water and ions distributions in
periodic systems; the code can compute accurate gradients that can
be used in minimizations or molecular dynamics simulations. The new
method includes an extension of the OZ equation needed to yield charge
neutrality for charged solutes which requires an additional contribution
to the excess chemical potential that has not been previously identified;
this is an important consideration for nucleic acids or any other
charged system where most or all of the counter- and co-ions are part
of the “disordered” solvent. We present of several calculations of
protein, RNA and small molecule crystals to show that X-ray scattering
intensities and solvent structure predicted by the periodic 3D-RISM
solvent model are in closer agreement with experiment than are intensities
computed using the default flat solvent model in the refmac5 or phenix
refinement programs, with the greatest improvement in the 2 to 4 Å
range. Prospects for incorporating integral equation models into crystallographic
refinement are discussed.
\end{abstract}
\maketitle

\section{Introduction\label{sec:Introduction}}

Ions and water molecules have been long known to play crucial roles
in governing biomolecular stability and function. Elucidating how
ions and water molecules distribute themselves around the solutes
should provide valuable insights into how those molecules function,
and also provide experimental tests for theoretical predictions. However,
there are few methods that directly probe the distributions of ions
and water molecules around macromolecules. In solution, excess numbers
of waters and ions around a macromolecule can be obtained using atomic
emission spectroscopy,\citep{Bai07,Gebala_JAmChemSoc_2015_v137_p14705}
small-angle X-ray scattering,\citep{Pabit09,Pabit10,Meisburger15,Nguyen16c}
or measurements of partial molar volumes.\citep{Chalikian03a,Chalikian07,Chalikian08,Son14a}
These techniques, however, give relatively little information about
the distribution of water and ions in the vicinity of a biomolecule.

In principle, much more detailed information is available from X-ray
diffraction studies on biomolecular crystals, and it is common to
include some number of ``bound'' (or localized) water molecules
and ions in a refined atomic model that has been optimized to fit
observed scattering intensities. These locations are typically identified
as features in a difference electron density map that satisfy criteria
for both intensity (percent occupation) and geometry. However, the
``bound'' solvent molecules generally make up only a small fraction
of the total solvent; the remainder is typically modeled as a flat
distribution, usually with density and B-factor components that are
adjusted to optimize the fit of the total model to observed intensities.
The limitations of such a flat-density model are thought to contribute
to the ``R-factor gap'', which reflects the nearly universal observation
that differences between computed and observed intensities in macromolecular
crystallography are much greater than the experimental uncertainties,
prompting searches for better models.\citep{Holton14}

In this paper, we develop and apply a novel integral equation models
(3D-RISM) to predict the solvent distribution in both small molecule
and macro-molecular crystals of proteins and nucleic acids. We present
results from a newly-developed periodic version of the existing non-perioidic
3D-RISM models in Amber\citep{Luchko10,Luchko12a}. Particular attention
is paid to the way in which charged solutes are handled to ensure
electroneutrality of the entire unit cell, that is, to ensure that
the distribution of ions in the solvent counterbalances the net charge
of the solute. 3D-RISM has been used in non-periodic systems to predict
location of site bound water and ions as well as to quantities reporting
on the diffuse and territorial binding modes of solvent particles
(ion counting, scattering profiles) as well as to give quantitative
energetics of solvation or small molecule binding to biomolecules.\citep{Ratkova15,Kovalenko15,Giambasu14,Giambasu19,Sugita20}
Here we explore the application of similar ideas to crystalline systems.

\section{Reference Interaction Site Model for periodic systems \label{sec:RISM-periodic}}

The core principle of RISM is to find the single particle density
distributions that minimize the excess chemical potential in response
to an external potential arising from a molecular solute. The basic
idea, and the approximations involved, have been discussed many times,\citep{Luchko10,Ratkova15}
and we only give a brief summary here. In principle, the distribution
of solvent molecules around a (fixed) solute is a six-dimensional
quantity, describing the translation and orientations of the solvent
molecules. The 3D-RISM formalism reduces these to three-dimensions
by decomposing polyatomic solvents (such as water molecules) into
atomic contributions, such that the resulting solvent density distributions
contain only a spatial dependence, $\rho_{\gamma}\left(\mathbf{r}\right)$,
and can be represented by scalar densities on 3D grids. Here, the
solvent index $\gamma$ would range over H and O sites in water, and
over mobile atomic cations such as Na\textsuperscript{+} and Cl\textsuperscript{-}.

An Ornstein-Zernike-like equation relates the total correlation function,
$h_{\gamma}\left(\mathbf{r}\right)=g_{\gamma}(\mathbf{r})-1$, and
direct correlation function, $c_{\gamma}\left(\mathbf{r}\right)$,
through a convolution (denoted by $*$):
\begin{equation}
h_{\gamma}^{\text{OZ}}\left(\mathbf{r}\right)=\sum_{\alpha}c_{\alpha}\left(\mathbf{r}\right)*\chi_{\alpha\gamma}\left(r\right)\label{eq:OZequation}
\end{equation}
Here, $\chi_{\alpha\gamma}\left(r\right)=\omega_{\alpha\gamma}\left(r\right)+\rho_{\alpha}h_{\alpha\gamma}\left(r\right)$
is the site-site solvent-susceptibility of solvent sites $\alpha$
and $\gamma$ and describes the orientationally averaged bulk properties
of the solvent, where $\omega_{\alpha\gamma}\left(\mathbf{r}\right)$
is an intramolecular correlation matrix, $\rho_{\alpha}$ is the bulk
number density, and $h_{\alpha\gamma}\left(r\right)$ is the total
correlation function. These values are pre-computed (generally by
a ``1D-RISM'' approach) for the reference solvent using the dielectrically
consistent RISM (DRISM) integral equation~\citep{Perkyns_ChemPhysLett_1992_v190_p626,Perkyns_JChemPhys_1992_v97_p7656}.
As in earlier work,\citep{Luchko10,Luchko12a} entities with two subscripts,
such as $h_{\alpha\gamma}\left(r\right)$ , refer to solvent--solvent
interactions, whereas a single subscript, such as $h_{\gamma}^{OZ}\left(\mathbf{r}\right)$,
refers to solvent site $\gamma$ at point $\mathbf{r}$ on the three-dimensional
grid surrounding the solute.

Eq. \ref{eq:OZequation} is augmented by a 3D closure relation: 
\begin{multline}
h_{\gamma}^{\text{closure}}\left(\mathbf{r}\right)\\
=\exp\left\{ -\beta u_{\gamma}\left(\mathbf{r}\right)+h_{\gamma}^{\text{OZ}}\left(\mathbf{r}\right)-c_{\gamma}\left(\mathbf{r}\right)+b_{\gamma}\left(\mathbf{r}\right)\right\} -1\label{eq:ClosureEquation}
\end{multline}
where $b_{\gamma}\left(\mathbf{r}\right)$ is the bridge function,
which is only known as an infinite series of functionals and is always
subject to some approximation\citep{Hansen_Book_TheoryofSimpleLiquidswithApplicationsofSoftMatter_2013}.
Among the many closure relations that have been developed, in this
work we use family of closures related to the hypernetted chain (HNC)
closure \citep{Morita_ProgTheorPhys_1958_v20_p920} where the bridge
function is simply set to zero. HNC produces good results for ionic~\citep{Howard_JPhysChemB_2011_v115_p547,Rasaiah_JChemPhys_1972_v56_p248,Hansen_PhysRevA_1975_v11_p2111}
and polar systems~\citep{Hirata_ChemPhysLett_1981_v83_p329,Hirata_JChemPhys_1982_v77_p509}
and has an exact, closed form expression for the excess chemical potential
\citep{Singer_MolPhys_2006_v55_p621}. Since HNC solutions are often
difficult to converge one can use an intermediaries such as the so-called
partial series expansion of order-$n$ (PSE-$n$) \citep{Kast_JChemPhys_2008_v129_p236101}
of HNC as a Taylor series expansion when the exponent in Eq. \ref{eq:ClosureEquation}
is positive:

\begin{equation}
h_{\gamma}^{\text{PSE}-n}\left(\mathbf{r}\right)=\left\{ \begin{aligned}\exp\left\{ t_{\gamma}\left(\mathbf{r}\right)\right\} -1 & \qquad t_{\gamma}\left(\mathbf{r}\right)<0\\
\sum_{i=1}^{n}\frac{{t_{\gamma}\left(\mathbf{r}\right)}^{i}}{i!} & \qquad t_{\gamma}\left(\mathbf{r}\right)\ge0
\end{aligned}
\right.\label{eq:PSE-closure}
\end{equation}
\[
t_{\gamma}\left(\mathbf{r}\right)=-\beta u_{\gamma}\left(\mathbf{r}\right)+h_{\gamma}^{\text{OZ}}\left(\mathbf{r}\right)-c_{\gamma}\left(\mathbf{r}\right).
\]
where HNC is the limiting case as $n\rightarrow\infty$. As for HNC,
the PSE-$n$ family of closures have an exact, closed form expression
for the chemical potential. The form of this approximation has a major
impact on the convergence of calculations as well as on resulting
thermodynamic quantities and correlation functions.

The goal of the self-consistent 3D-RISM procedure can be viewed as
finding a direct correlation function $c_{\gamma}\left(\mathbf{r}\right)$
such that $h_{\gamma}^{\text{OZ}}$ and $h_{\gamma}^{\text{closure}}$
become identical at all grid points to within some (fairly tight)
tolerance. In existing, non-periodic, implementations, the convolution
required in Eq. \ref{eq:OZequation} is carried out via fast Fourier
transforms in a rectangular box surrounding the solute, and additional
terms that account for solvent outside of the artificial box are added
to this. Key differences are that the electrostatic and Lennard-Jones
potentials that appear in Eq. \ref{eq:PSE-closure} need to take periodic
boundary conditions into account and that some special considerations
are needed, when the solute has a net charge, to ensure charge neutrality
for each unit cell. While periodic methods (e.g., particle mesh Ewald
(PME) and Ewald summation) have been used before to synthesize the
long-range electrostatic potential on a 3D grid, these approaches
assume infinite dilution of the solute and employ corrections to capture
the long-range behavior of the solvent when calculating the excess
chemical potential.\citep{Kovalenko99,Heil15} In contrast, we use
periodic boundary conditions throughout the method described in the
next two sections.

\subsection{Constructing the periodic solute potential.}

The closure functional equation requires the mapping of the solute
potential onto regular grids that covers the entire unit cell with
one potential grid for each type of solvent site encompassing both
Lennard-Jones and electrostatic components. Mapping the electrostatic
potential follows the smooth PME procedure used in molecular dynamics
simulations \citep{Darden93,Essmann95} although the grid spacing
is smaller, typically 0.5 Å. Lennard-Jones interactions between solute
atoms and all solvent types are calculated at each grid point using
a distance cutoff (default is 9 Å) and the minimum-image convention.
The same convention is used for the short-range part of the electrostatic
potential, where the bare Coulomb interaction is replaced by $\mathrm{erfc}\left(\beta\left|\mathbf{r}-\mathbf{r}_{i}\right|\right)/\left|\mathbf{r}-\mathbf{r}_{i}\right|$
where $\mathbf{r}$ is the position of a solute atom, and $\mathbf{r}_{i}$
a point on the grid. The remaining, long-range part of the periodic
Coulomb potential is solved for in the reciprocal space, via fast
Fourier transforms (FFT) and follows these steps: :\citep{Darden93,Essmann95}
\begin{enumerate}
\item Interpolate the solute atomic charges to the direct space Cartesian
grid. The current version of the code relies on the smooth PME (SPME)
approach, which uses a cardinal b-spline of order 4 or 6 to interpolate
the source charge to the grid. The b-spline interpolation has a roughly
Gaussian character at high polynomial orders, and has the desirable
trait that integration of its weights over the region of interpolation
equals unity.
\item Convert the source charge grid from real space to reciprocal space
using an FFT.
\item Compute the electrostatic potential and spatial derivatives (electrostatic
field) on the grid using a convolution with a reciprocal space representation
of the Gaussian kernel and its derivatives; in reciprocal space the
convolution is a simple multiplication, and the electrostatic interaction
potential Green's function is $k^{-2}$ .
\item Obtain the real space representation of the electrostatic potential
and electrostatic field using an inverse FFT.
\end{enumerate}
Full details of this procedure are given elsewhere.\citep{Johnson16a}

\subsection{Solving the 3D-RISM equations.\label{subsec:neutralize}}

As noted above, solving the 3D-RISM equations amounts to finding a
direct correlation functional, $c_{\gamma}$, for each solute site
$\gamma$ that minimizes the residual: $\Delta c_{\gamma}\left(\mathbf{r}\right)\equiv h_{\gamma}^{\text{closure}}\left(\mathbf{r}\right)-h_{\gamma}^{\text{O}Z}\left(\mathbf{r}\right)$
for a specific iteration. Calculations are initialized with a guess
for each $c_{\gamma}$, which are chosen to be uniformly zero, although
the code allows for a user-provided starting point which can accelerate
convergence for systems difficult to solve. Each self-consistent cycle
begins with computing $h_{\gamma}^{\text{OZ}}$ in the reciprocal
space using Eq. \ref{eq:OZequation}, followed by a switch to the
real space, where $h_{\gamma}^{\text{closure}}$ is computed using
Eq. \ref{eq:ClosureEquation}, and ends by modifying the current guess
for $c_{\gamma}$ using the modified direct inversion of the iterative
subspace (MDIIS) procedure\citep{Kovalenko00a,Luchko10} based upon
$\Delta c_{\gamma}$ and a specified number of past $c_{\gamma}$
solutions. This cycle is repeated until the root-mean squared residual,
$\text{RMS}\left(\Delta c_{\gamma}\right)$, reaches a pre-determined
threshold, which is typically 10\textsuperscript{-10} if gradients
are needed (such as in the case of minimization or dynamics), and
10\textsuperscript{-6} if one just needs thermodynamic parameters
or solvent distribution functions. Once convergence is obtained, there
is no longer any distinction between $h^{\text{closure}}$ and $h^{\text{OZ}}$.

This procedure is complicated when charged solutes are used: here
one wants the solute net charge to be neutralized by the converged
ion distribution of the solvent. However, a consequence of using PME
is that a uniform neutralizing background charge is imposed on the
system; i.e., the effective net charge of the solute is always zero
if only the PME component of the potential is used. As a result, the
$h^{\text{OZ}}$ distribution arising from Eq. \ref{eq:OZequation}
will also be neutral, which is a problem when the solute charge is
non-zero. We describe here a procedure modeled after that used by
Kovalenko and Hirata for non-periodic 3D-RISM\citep{Kovalenko99,Kovalenko00a},
which modifies the OZ direct correlation function to account for this
implicit background charge. We first note the potential energy due
to the solvent site $\gamma$ interacting with a non-neutral solute
is
\begin{equation}
u_{\gamma}\left(\mathbf{r}\right)=u_{\gamma}^{\text{PME}}\left(\mathbf{r}\right)-u_{\gamma}^{\text{bk}}\left(\mathbf{r}\right)=u_{\gamma}^{\text{PME}}\left(\mathbf{r}\right)-q_{\gamma}\phi^{\text{bk}}\left(\mathbf{r}\right),\label{eq:full-potential-energy}
\end{equation}
where $u_{\gamma}^{\text{PME}}\left(\mathbf{r}\right)$ is the potential
energy calculated by PME, $u_{\gamma}^{\text{bk}}\left(\mathbf{r}\right)$,
is the potential energy due to the neutralizing background charge
and $\phi^{\text{bk}}\left(\mathbf{r}\right)$ is the background potential
imposed by PME. Since $u_{\gamma}^{\text{bk}}$ represents the interaction
of the solvent charge with an infinite background charge density,
it diverges, and we cannot directly use Eq. \ref{eq:full-potential-energy}
in Eq. \ref{eq:ClosureEquation} as it stands. However, an analytic
expression for $\phi^{\text{bk}}\left(\mathbf{r}\right)$ can be found
in reciprocal space: using the fact that the background charge distribution
is $q^{\text{bk}}\left(\mathbf{r}\right)=-Q_{\text{solute}}/V_{\text{cell}}$,
we can write Poisson's equation as
\begin{equation}
\hat{\phi}^{\text{bk}}\left(\mathbf{k}\right)=4\pi\frac{\hat{q}^{\text{bk}}\left(\mathbf{k}\right)}{k^{2}}=-\delta\left(k\right)\frac{Q_{\text{solute}}}{V_{\text{cell}}}\frac{4\pi}{k^{2}}.\label{eq:phibk}
\end{equation}
The restriction to $k=0$ yields a uniformly distributed quantity
in real-space but has the expected singularity at $k=0$. Using HNC
for simplicity, Eq. \ref{eq:ClosureEquation} can then be written
as
\begin{align}
h_{\gamma}^{\text{HNC}}(\mathbf{r})+1 & =\exp\left[-\beta u_{\gamma}(\mathbf{r})+h_{\gamma}^{\text{OZ}}(\mathbf{r})-c_{\gamma}(\mathbf{r})\right]\nonumber \\
 & =\exp\left[-\beta u_{\gamma}^{\text{PME}}(\mathbf{r})+h_{\gamma}^{\text{OZ}}(\mathbf{r})-\tilde{c}_{\gamma}(\mathbf{r})\right],\label{eq:hclosuretilde}
\end{align}
where we have grouped the background charge contribution with $c_{\gamma}\left(\mathbf{r}\right)$
to define a renormalized direct correlation function
\begin{equation}
\tilde{c}_{\gamma}\left(\mathbf{r}\right)=c_{\gamma}\left(\mathbf{r}\right)-\beta q_{\gamma}\phi^{\text{bk}}\left(\mathbf{r}\right).\label{eq:ctildedef}
\end{equation}
The Ornstein-Zernike equation, Eq. \ref{eq:OZequation}, is then 
\begin{align}
h_{\gamma}^{\text{OZ}}\left(\mathbf{r}\right) & =\sum_{\alpha}c_{\alpha}\left(\mathbf{r}\right)*\chi_{\alpha\gamma}\left(r\right)\nonumber \\
 & =\sum_{\alpha}\left[\tilde{c}_{\alpha}\left(\mathbf{r}\right)*\chi_{\alpha\gamma}\left(r\right)+\beta u_{\alpha}^{\text{bk}}\left(\mathbf{r}\right)*\chi_{\alpha\gamma}\left(r\right)\right].\label{eq:hoztilde}
\end{align}
Taking Fourier transforms, which we denote by $\hat{\cdot}$:
\begin{align}
\sum_{\alpha}\beta\hat{u}_{\alpha}^{\text{bk}}\left(\mathbf{k}\right)\hat{\chi}_{\alpha\gamma} & \left(k\right)\nonumber \\
 & =-\sum_{\alpha}\beta\delta\left(k\right)q_{\alpha}\frac{Q_{\text{solute}}}{V_{\text{cell}}}\frac{4\pi}{k^{2}}\hat{\chi}_{\alpha\gamma}\left(k\right)\nonumber \\
 & =-4\pi\beta\frac{Q_{\text{solute}}}{V_{\text{cell}}}\delta\left(k\right)\lim_{k\rightarrow0}\sum_{\alpha}\frac{q_{\alpha}}{k^{2}}\hat{\chi}_{\alpha\gamma}\left(k\right)\nonumber \\
 & \equiv\hat{h}_{\gamma}^{\text{bk}}\delta\left(k\right)\label{eq:hbk}
\end{align}
This depends only on $Q_{\text{solvent}}$, $V_{\text{cell}},$ and
properties of the bulk solvent and evaluates to a constant when going
back to real space. Even though $\hat{\phi}^{\text{bk}}\left(\mathbf{k}\right)$
in Eq. \ref{eq:phibk} has a singularity at $k=0$, $\hat{h}_{\gamma}^{\text{bk}}$in
Eq. \ref{eq:hbk} is finite. In practice, we use a polynomial interpolation
procedure based on Neville's algorithm to numerically extrapolate
values at finite $k$ in Eq. \ref{eq:hbk} to the $k=0$ limit.

Modifying $h^{\text{OZ}}(\boldsymbol{r})$, by a constant would seem
yield a distribution function $g(\boldsymbol{r})\equiv h(\boldsymbol{r})+1$
that is not zero inside solute atoms. But during the self-consistent
cycle this shift is immediately followed by an application of the
closure relation, Eq. \ref{eq:hclosuretilde}, with a contribution
$\exp[-\beta u(\boldsymbol{r})]$ that serves to prevent solvent species
from being close to solute atoms, as discussed in Refs. \citealp{Kovalenko99,Kovalenko00a}.
Solving Eqs. \ref{eq:hclosuretilde} and \ref{eq:hoztilde}, rather
than the original Eqs. \ref{eq:OZequation} and \ref{eq:ClosureEquation},
implies that the renormalized $\tilde{c}_{\gamma}\left(\mathbf{r}\right)$
is used throughout the algorithm in Fig. \ref{fig:algorithm}. By
doing so, the solvent distribution will exactly neutralize the solute
charge even though we only use the neutralized potential energy, $u_{\gamma}^{\text{PME}}\left(\mathbf{r}\right)$.

In the end, the procedure for charged solutes is only slightly modified
from that used for neutral solutes: we use $\tilde{c}$ rather than
$c$ (Eq. \ref{eq:ctildedef}) and ``shift'' $h_{\gamma}^{\text{OZ}}$
by $h_{\gamma}^{\text{bk}}$ (Eq. \ref{eq:hbk}). Pseudo-code for
this process is given in Fig. \ref{fig:algorithm}.

\begin{center}
\begin{figure}[t]
\begin{centering}
\includegraphics{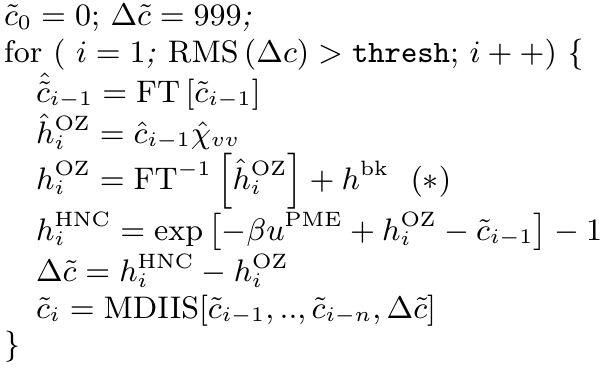}
\par\end{centering}
\caption{\label{fig:algorithm}Pseudo-code for an algorithm to solve the 3D-RISM
equations using $\tilde{c}$ and a \textquotedblleft shift\textquotedblright{}
in $h^{\text{OZ}}$ (starred line). The code assumes HNC closure for
simplicity; other starting estimates for $\tilde{c}_{0}$ may be used,
such as the result from a previous step of minimization or MD. $\text{FT}$
is Fourier transform; $\text{MDIIS}$ is a version of the direct inversion
of the iterative subspace;\citep{Kovalenko00a,Luchko10} and \texttt{thresh}
is a convergence threshold. For a neutral solute, or for pure water
as a solvent, $h^{\text{bk}}$ is zero.}
\end{figure}
\par\end{center}

\subsection{Computing the excess chemical potential for a shifted OZ model\label{subsec:Computing-the-excess}}

When the algorithm in Figure \ref{fig:algorithm} is complete, we
have the total correlation function for the charged solute, $h_{\gamma}\left(\mathbf{r}\right)$,
but the direct correlation function used in the calculation , $\tilde{c}_{\gamma}\left(\mathbf{r}\right)$,
contains the background potential energy. This must be accounted for
when calculating the excess chemical potential. Here, we follow Kovalenko
and Hirata \citep{Kovalenko99} and start by considering the Kirkwood
charging formula for the excess chemical potential, but we use $u_{\gamma}^{\text{PME}}\left(\mathbf{r},\lambda\right)$,
rather than the full potential, 
\begin{align}
\Delta\mu & =\sum_{\gamma}\rho_{\gamma}\int_{0}^{1}\int_{V_{\text{cell}}}\frac{\partial u_{\gamma}^{\text{PME}}\left(\mathbf{r},\lambda\right)}{\partial\lambda}g_{\gamma}^{\text{}}\left(\mathbf{r},\lambda\right)\,d\mathbf{r}d\lambda,\label{eq:kirkwood-charging-1}
\end{align}
where $\lambda$ is a coupling parameter between the solute and solvent.
To find a closed form expression for $\Delta\mu$, we will recast
the integrand in the form of an exact differential in $\lambda$.
To begin, we consider the functional variation of the HNC closure,
Eq. \ref{eq:hclosuretilde}:
\begin{multline}
\delta h_{\gamma}\left(\mathbf{r}\right)=-g_{\gamma}\left(\mathbf{r}\right)\beta\delta u_{\gamma}^{\text{PME}}\left(\mathbf{r}\right)+g_{\gamma}\left(\mathbf{r}\right)\delta h_{\gamma}\left(\mathbf{r}\right)\\
-g_{\gamma}\left(\mathbf{r}\right)\delta\tilde{c}_{\gamma}\left(\mathbf{r}\right),\label{eq:var-HNC}
\end{multline}
which is valid for any variation, including variation with respect
to $\lambda$. We can solve for the integrand in Eq. \ref{eq:kirkwood-charging-1}
using $g_{\gamma}=h_{\gamma}+1$ and $\delta\left[h_{\gamma}^{2}\left(\mathbf{r},\lambda\right)\right]=2h_{\gamma}\left(\mathbf{r},\lambda\right)\delta h_{\gamma}\left(\mathbf{r},\lambda\right)$
to arrive at
\begin{equation}
\beta g_{\gamma}\left(\mathbf{r}\right)\delta u_{\gamma}^{\text{PME}}\left(\mathbf{r}\right)=\delta\left[\frac{\left(h_{\gamma}\left(\mathbf{r}\right)\right)^{2}}{2}-\tilde{c}_{\gamma}\left(\mathbf{r}\right)\right]-h_{\gamma}\left(\mathbf{r}\right)\delta c_{\gamma}\left(\mathbf{r}\right).\label{eq:kirkwood-charging-integrand-1-1}
\end{equation}
The first term is already in the form of an exact differential, and
we use the following variation to cast the second term to a useful
form:
\begin{align*}
\delta\sum_{\gamma}\rho_{\gamma} & \int_{V_{\text{cell}}}h_{\gamma}\left(\mathbf{r}\right)\tilde{c}_{\gamma}\left(\mathbf{r}\right)\,d\mathbf{r}\\
 & =\int_{V_{\text{cell}}}\sum_{\gamma}\rho_{\gamma}\sum_{\alpha}\Bigl[\tilde{c}_{\alpha}\left(\mathbf{r}\right)*\chi_{\alpha\gamma}\left(r-r'\right)\delta\tilde{c}_{\gamma}\left(\mathbf{r}\right)\\
 & \phantom{=\int_{V_{\text{cell}}}\sum_{\gamma}\rho_{\gamma}\sum_{\alpha}}+\tilde{c}_{\gamma}\left(\mathbf{r}\right)\delta\tilde{c}_{\alpha}\left(\mathbf{r}\right)*\chi_{\alpha\gamma}\left(r-r'\right)\Bigr]\\
 & \phantom{=\int_{V_{\text{cell}}}}-\sum_{\gamma}\rho_{\gamma}h_{\gamma}^{\text{bk}}\delta\tilde{c}_{\gamma}\left(\mathbf{r}\right)\,d\mathbf{r},
\end{align*}
where we have substituted $h_{\gamma}$using Eq. \ref{eq:OZequation}.
As $\chi_{\alpha\gamma}\left(r-r'\right)$ is a property of the bulk
solvent and, consequently, invariant,
\begin{align*}
\delta\sum_{\gamma}\rho_{\gamma}\int_{V_{\text{cell}}}h_{\gamma}\left(\mathbf{r}\right) & \tilde{c}_{\gamma}\left(\mathbf{r}\right)\,d\mathbf{r}\\
 & =\int_{V_{\text{cell}}}\sum_{\gamma}\rho_{\gamma}2h_{\gamma}\left(\mathbf{r}\right)\delta\tilde{c}_{\gamma}\left(\mathbf{r}\right)\\
 & \phantom{=\int_{V_{\text{cell}}}}+\sum_{\gamma}\rho_{\gamma}h_{\gamma}^{\text{bk}}\delta\tilde{c}_{\gamma}\left(\mathbf{r}\right)\,d\mathbf{r}.
\end{align*}
Upon simple rearrangement, we can write 
\begin{multline}
\sum_{\gamma}\rho_{\gamma}\int_{V_{\text{cell}}}h_{\gamma}\left(\mathbf{r}\right)\delta\tilde{c}_{\gamma}\left(\mathbf{r}\right)\,d\mathbf{r}\\
=\frac{1}{2}\sum_{\gamma}\rho_{\gamma}\int_{V_{\text{cell}}}\delta\left[h_{\gamma}\left(\mathbf{r}\right)\tilde{c}_{\gamma}\left(\mathbf{r}\right)-h_{\gamma}^{\text{bk}}\tilde{c}_{\gamma}\left(\mathbf{r}\right)\right]\,d\mathbf{r},\label{eq:kirkwood-hc-exact-1}
\end{multline}
for which the left hand side now has the form of an exact differential.

Taking together Eqs. \ref{eq:kirkwood-charging-1}, \ref{eq:kirkwood-charging-integrand-1-1}
and \ref{eq:kirkwood-hc-exact-1} we can derive a final expression
for the excess chemical potential,
\begin{align}
\Delta\mu^{\text{HNC}} & =\beta^{-1}\sum_{\gamma}\rho_{\gamma}\int_{V_{\text{cell}}}\frac{h_{\gamma}^{2}\left(\mathbf{r}\right)}{2}-\left(1-\frac{h_{\gamma}^{\text{bk}}}{2}\right)\tilde{c}_{\gamma}\left(\mathbf{r}\right)\nonumber \\
 & \phantom{=kT\sum_{\gamma}\rho_{\gamma}\int_{V_{\text{cell}}}}-\frac{h_{\gamma}\left(\mathbf{r}\right)\tilde{c}_{\gamma}\left(\mathbf{r}\right)}{2}\,d\mathbf{r},\label{eq:exchem-hnc-shifted}
\end{align}
which is identical to the HNC expression for Eq. \ref{eq:OZequation}
except for the use of $\tilde{c}_{\gamma}\left(\mathbf{r}\right)$
and an additional term, $-kT\sum_{\gamma}\rho_{\gamma}\int_{V_{\text{cell}}}-\frac{1}{2}h_{\gamma}^{\text{bk}}\tilde{c}_{\gamma}\,d\mathbf{r}$,
resulting from using the renormalized direct correlation function,
$\tilde{c}_{\gamma}\left(\mathbf{r}\right)$. This additional term
will be present for all closures with a closed form expression of
the excess chemical potential. A similar treatment for the PSE-$n$
family of closures \citep{Kast_JChemPhys_2008_v129_p236101}, which
includes the Kovalenko-Hirata (KH) closure \citep{Kovalenko00a} is
presented in Appendix 1.

\subsection{Computing solvation forces on the periodic solute atoms}

A closed form expression for the solvation force on atom $i$ due
to Eq. \ref{eq:exchem-hnc-shifted},
\[
\mathbf{f}_{i}\left(\text{\ensuremath{\mathbf{R}_{i}}}\right)=\frac{\partial}{\partial\mathbf{R}_{i}}\Delta\mu,
\]
may also be derived following the approach of Kovalenko and Hirata
\citep{Kovalenko99}. For simplicity, we will again use the HNC expression
for the excess chemical, Eq. \ref{eq:exchem-hnc-shifted}, as the
approach is easily extended to any closure with a closed form expression
for the excess chemical potential. For example, the variation of Eq.
\ref{eq:exchem-hnc-shifted} is given by
\begin{align}
\delta\Delta\mu^{\text{HNC}} & =kT\sum_{\gamma}\rho_{\gamma}\int_{V_{\text{cell}}}\biggl[h_{\gamma}\left(\mathbf{r}\right)\delta h_{\gamma}\left(\mathbf{r}\right)\nonumber \\
 & \phantom{=kT\sum_{\gamma}\rho_{\gamma}\int_{V_{\text{cell}}}}-\delta\left(\frac{h_{\gamma}\left(\mathbf{r}\right)\tilde{c}_{\gamma}\left(\mathbf{r}\right)}{2}\right)\\
 & \phantom{=kT\sum_{\gamma}\rho_{\gamma}\int_{V_{\text{cell}}}}-\left(1-\frac{h_{\gamma}^{\text{bk}}}{2}\right)\delta\tilde{c}_{\gamma}\left(\mathbf{r}\right)\,d\mathbf{r}\biggr].\label{eq:variation-exchem-hnc}
\end{align}
However, variations in the total and direct correlation functions
are difficult to numerically compute and we would like to confine
the variation to the potential only. Meanwhile, solving for $g_{\gamma}\left(\mathbf{r}\right)\beta\delta u_{\gamma}^{\text{\text{PME}}}\left(\mathbf{r}\right)$
in Eq. \ref{eq:var-HNC} and simplifying we have
\begin{multline*}
g_{\gamma}\left(\mathbf{r}\right)\beta\delta u_{\gamma}^{\text{PME}}\left(\mathbf{r}\right)\\
=h_{\gamma}\left(\mathbf{r}\right)\delta h_{\gamma}\left(\mathbf{r}\right)-h_{\gamma}\left(\mathbf{r}\right)\delta\tilde{c}_{\gamma}\left(\mathbf{r}\right)-\delta\tilde{c}_{\gamma}\left(\mathbf{r}\right).
\end{multline*}
Using Eq. \ref{eq:kirkwood-hc-exact-1}, we can write
\begin{align*}
\sum_{\gamma}\rho_{\gamma}\int_{V_{\text{cell}}} & g_{\gamma}\left(\mathbf{r}\right)\beta\delta u_{\gamma}^{\text{PME}}\left(\mathbf{r}\right)\,d\mathbf{r}\\
 & =\sum_{\gamma}\rho_{\gamma}\int_{V_{\text{cell}}}h_{\gamma}\left(\mathbf{r}\right)\delta h_{\gamma}\left(\mathbf{r}\right)\\
 & \phantom{=\sum_{\gamma}\rho_{\gamma}\int_{V_{\text{cell}}}}-\delta\left(\frac{h_{\gamma}\left(\mathbf{r}\right)\tilde{c}_{\gamma}\left(\mathbf{r}\right)}{2}\right)\\
 & \phantom{=\sum_{\gamma}\rho_{\gamma}\int_{V_{\text{cell}}}}-\left(1-\frac{h_{\gamma}^{\text{bk}}}{2}\right)\delta\tilde{c}_{\gamma}\left(\mathbf{r}\right)\,d\mathbf{r}.
\end{align*}
As the right hand side matches the summation in Eq. \ref{eq:variation-exchem-hnc},
we have
\[
\delta\Delta\mu^{\text{HNC}}=\sum_{\gamma}\rho_{\gamma}\int_{V_{\text{cell}}}g_{\gamma}^{\text{}}\left(\mathbf{r}\right)\delta u_{\gamma}^{\text{PME}}\left(\mathbf{r}\right)\,d\mathbf{r}.
\]
Taking the variation with respect to the position of a solute atom,
$\mathbf{R}_{i}$, we have
\begin{align}
\mathbf{f}_{i}\left(\text{\ensuremath{\mathbf{R}_{i}}}\right) & =\frac{\partial}{\partial\mathbf{R}_{i}}\Delta\mu^{\text{HNC}}\nonumber \\
 & =\sum_{\gamma}\rho_{\gamma}\int_{V_{\text{cell}}}g_{\gamma}\left(\mathbf{r}\right)\frac{\partial}{\partial\mathbf{R}_{i}}u_{\gamma}^{\text{PME}}\left(\mathbf{r}\right)\,d\mathbf{r}.\label{eq:gradients}
\end{align}
This expression is same as that for the standard 3D-RISM equation
and independent of $h_{\gamma}^{\text{bk}}$.

\section{Methods}

\begin{table*}
\begin{ruledtabular}
\centering{}%
\begin{tabular}{ldcc}
PDB/CSD ID & \multicolumn{1}{c}{Spacing} & Grid size & Solvent\tabularnewline
\hline 
ANOMEW\citep{Friscic11} & 0.33 & $84\times72\times96$ & 0.005M MgCl$_{2(aq)}$\tabularnewline
1AHO \citep{Smith97a} & 0.4 & $120\times108\times80$ & Water\tabularnewline
2IGD \citep{Derrick94} & 0.35 & $108\times120\times126$ & Water\tabularnewline
1BZR \citep{Kachlova99} & 0.35 & $108\times190\times192$ & Water\tabularnewline
4LZT \citep{Walsh98} & 0.35 & $80\times96\times108$ & Water\tabularnewline
2LZT \citep{ramanadham1990refinement} & 0.35 & $80\times96\times108$ & Water\tabularnewline
4YUL \citep{Keedy15} & 0.35 & $126\times160\times256$ & Water\tabularnewline
2A43 \citep{pallan2005crystal} & 0.35 & $160\times160\times160$ & 0.02M MgCl$_{2}$, 0.14M KCl$_{(aq)}$\tabularnewline
480D \citep{Correll99} & 0.35 & $90\times90\times224$ & 1M NaCl$_{(aq)}$\tabularnewline
2QUS \citep{chi2008capturing} & 0.35 & $80\times160\times210$ & 1M NaCl$_{(aq)}$\tabularnewline
1Y0Q \citep{golden2005crystal} & 1.0 & $96\times144\times224$ & 0.02M MgCl$_{2}$, 0.14M KCl$_{(aq)}$\tabularnewline
2OIU \citep{robertson2007structural} & 1.0 & $48\times112\times80$ & 0.1M MgCl$_{2}$, 1.29M NaCl$_{(aq)}$\tabularnewline
\end{tabular}\caption{3D-RISM parameters for crystal structure optimization and energy minimization.
Grid spacing in Å. \label{tab:RISM-parameters}}
\end{ruledtabular}

\end{table*}

\subsection{Solute preparation}

With the exception of the heme group for myoglobin (PDB ID 1BZR) and
GTP in the hammerhead ribozyme (PDB ID 2QUS), all solvent and non-standard
residues were removed from the deposited crystal structures. All protein
and RNA structures were parameterized with the standard Amber charges
and Lennard-Jones parameters,\citep{Cornell95} which have not changed
since 1995. Naproxen was parameterized with the general Amber force
field 2 (GAFF2) \citep{Wang04c}. Parameters for hemoglobin\citep{Giammona84}
and GTP\citep{Meagher03} were taken from the Amber contributed parameter
database. The minimizations for 2OIU and 1Y0Q used the RNA ff99OL3
force field.\citep{Perez07,Zgarbova11}

\subsection{Solvent preparation}

Properties of the bulk solvent, including $\hat{\chi}_{\alpha\gamma}\left(k\right)$,
required for Eq. \ref{eq:hoztilde} were precomputed with \emph{rism1d}
from the AmberTools 21 \citep{Luchko10,case2021amber2021}. In all
cases, dielectrically consistent RISM (DRISM) \citep{Perkyns_JChemPhys_1992_v97_p7656}
was solved at a temperature of 298 K with a dielectric constant of
78.497 and the KH closure \citep{Kovalenko00a} on a grid with 0.025
Å spacing and 32768 points to a residual tolerance of $10^{-12}$.
The coincident SPC/E (cSPC/E) water model was used with Joung-Cheatham
parameters for monovalent ions \citep{Joung08} and Li-Merz 12-6 parameters
for divalent ions \citep{Li14}. Details of the solvent composition
for each solute can be found in Table \ref{tab:RISM-parameters}.

\subsection{3D-RISM calculations}

Eq. \ref{eq:hoztilde} was solved using \emph{sander} from AmberTools
21, modified as described in section \ref{sec:RISM-periodic}. Except
where described in Results, grid sizes and spacings are as detailed
in Table \ref{tab:RISM-parameters}. Calculations of solvation forces
(sections \ref{subsec:gradients} and \ref{subsec:implict-solvent})
were solved to a residual tolerance of $10^{-10}$, while all other
calculations were solved to a residual tolerance of $10^{-7}$. For
biomolecular crystals grid dimensions were selected to match the unit
cells of the deposited structures, with exceptions noted for the calculations
discussed in sections \ref{subsec:gradients} and \ref{subsec:Extrapolation}.
For the small molecule crystal naproxen calculation the original unit
cell was expanded 3, 7 and 3 times respectively along the three crystal
lattice vectors.

\section{Results}

We have applied this periodic 3D-RISM model to a variety of protein
and nucleic acid crystals. We begin with discussions of the accuracy
of forces on solute atoms arising from the gradients of the excess
chemical potential (Section \ref{subsec:gradients}), then look at
the way a periodic system extrapolates to a non-periodic limit as
the size of the periodic box surrounding a single solvent molecule
increases (Section \ref{subsec:Extrapolation}). These help to provide
confidence in the correctness of our implementation. We then look
at examples of the solvent distributions in biomolecules, comparing
to Xray scattering factors (Section \ref{subsec:xray}), and give
examples of predictions for electrostatic screening effects in RNA
crystals (Section \ref{subsec:implict-solvent}). These shows promising
results, but it is clear than many more studies will be needed to
map out the expected level of accuracy of this approach.

\subsection{Accuracy of atomic forces\label{subsec:gradients}}

The use of 3D-RISM as an implicit solvent requires accurate and rapid
calculation of atomic forces. Both speed and accuracy may depend upon
the system. Table \ref{tab:gradient-accuracy-1} gives some results
for a small RNA unit cell, with 108 nucleotides in four chains. We
compare gradients computed via Eq. \ref{eq:gradients} to those computed
with finite differences using Eq. \ref{eq:exchem-hnc-shifted}. There
is smooth convergence with respect to grid spacing for both $\Delta\mu$
and for the accuracy of the gradients, but very large grids can be
expensive. For the practical examples discussed below in Section \ref{subsec:implict-solvent}
we find that a 0.5 Å grid spacing gives results that hardly differ
from tighter grids. This is supported by the numbers of excess water
and ions presented in Table \ref{tab:gradient-accuracy-1}, which
show that key properties of the solvent distribution are converged
even at the larger grid spacings. The actual value of $\Delta\mu$
is not available from experiment, so grid artifacts in estimating
its value are of little consequence provided that the gradients and
solvent distributions are accurate. This appears to be the case for
even the largest grid spacings shown in the Table.

It is worth noting that the ``additional'' background contribution
of $-kT\sum_{\gamma}\rho_{\gamma}\int_{V_{\text{cell}}}-\frac{1}{2}h_{\gamma}^{\text{bk}}\tilde{c}_{\gamma}\,d\mathbf{r}$
in Eqs. \ref{eq:exchem-hnc-shifted} and \ref{eq:exchem-psen-shifted}
is key for periodic calculations. If this contribution is omitted,
the value of $\Delta\mu$ changes to $-468$ kcal/mol (for a 0.5 Å
grid spacing), and the mean and maximum absolute derivative errors
are 0.47 and 1.86 kcal/mol-Å, more than two orders of magnitude larger
than the values shown in Table \ref{tab:gradient-accuracy-1}. 

By comparison, for a single solute in a large box, this ``additional''
term is quite small. As an example, consider one chain of sarcin-ricin
from Table \ref{tab:gradient-accuracy-1}. Even with a fairly large
solute charge of $-26$, embedding this in a 120 Å box yields $\Delta\mu$
of -5941.41 kcal/mol without the ``correction'', and -5941.62 with
it, for a difference of 0.21 kcal/mol.

\begin{table*}
\begin{ruledtabular}
\begin{centering}
\begin{tabular}{dddddddd}
\multicolumn{1}{c}{Grid spacing} & \multicolumn{1}{c}{MAE} & \multicolumn{1}{c}{max} & \multicolumn{1}{c}{$\Delta\mu$} & \multicolumn{1}{c}{H\textsubscript{2}O} & \multicolumn{1}{c}{Mg\textsuperscript{2+}} & \multicolumn{1}{c}{K\textsuperscript{+}} & \multicolumn{1}{c}{Cl\textsuperscript{-}}\tabularnewline
\hline 
0.75 & 0.0053 & 0.0332 & 82.7 & -1049.6 & 19.89 & 56.79 & -7.43\tabularnewline
0.50 & 0.0026 & 0.0128 & 54.2 & -1047.6 & 19.89 & 56.79 & -7.43\tabularnewline
0.25 & 0.0006 & 0.0040 & 43.9 & -1047.3 & 19.89 & 56.79 & -7.43\tabularnewline
0.15 & 0.0004 & 0.0018 & 43.6 & -1047.2 & 19.89 & 56.79 & -7.43\tabularnewline
\end{tabular}
\par\end{centering}
\end{ruledtabular}

\caption{\label{tab:gradient-accuracy-1}Comparison of gradients computed via
Eq. \ref{eq:exchem-psen-shifted} to those computed with finite differences
using a displacement of 10\protect\textsuperscript{-4} Å. Grid spacing
is in Å. MAE is the mean absolute error, max is the maximum absolute
error (both in kcal/mol-Å) for the \emph{x,y,z} components of the
gradient for 20 randomly selected atoms. $\Delta\mu$ is the excess
chemical potential in kcal/mol. The final four columns give the excess
number of waters and ions. The system is one unit-cell of the sarcin-ricin
system PDB ID 480d, with 108 nucleotides and a solute charge of -104.
The solvent is 0.02 M MgCl\protect\textsubscript{2} plus 0.14M KCl
in water.}
\end{table*}

\subsection{Extrapolation to the infinite dilution regime\label{subsec:Extrapolation} }

The examples discussed above dealt with molecular crystals, where
solute molecules are in contact with their images in neighboring unit
cells, and the solvent volume is fairly small. Another application
might be to a single (dilute) solute surrounded by a buffer of solvent.
As the size of the unit cell increases, such a calculation should
approach the infinite dilution, non-periodic limit that has traditionally
been assumed in 3D-RISM applications. As noted above, these traditional
calculations actually employ a regular periodic grid in the vicinity
of the solute (to enable convolutions to be carried out via fast Fourier
Transforms), and add in estimates of the ``asymptotic'' contributions
from solvent outside the grid. Here we study the box-size dependence
of periodic 3D-RISM calculations that have a single solute molecule
at the origin.

\begin{figure}[t]
\centering{}\includegraphics{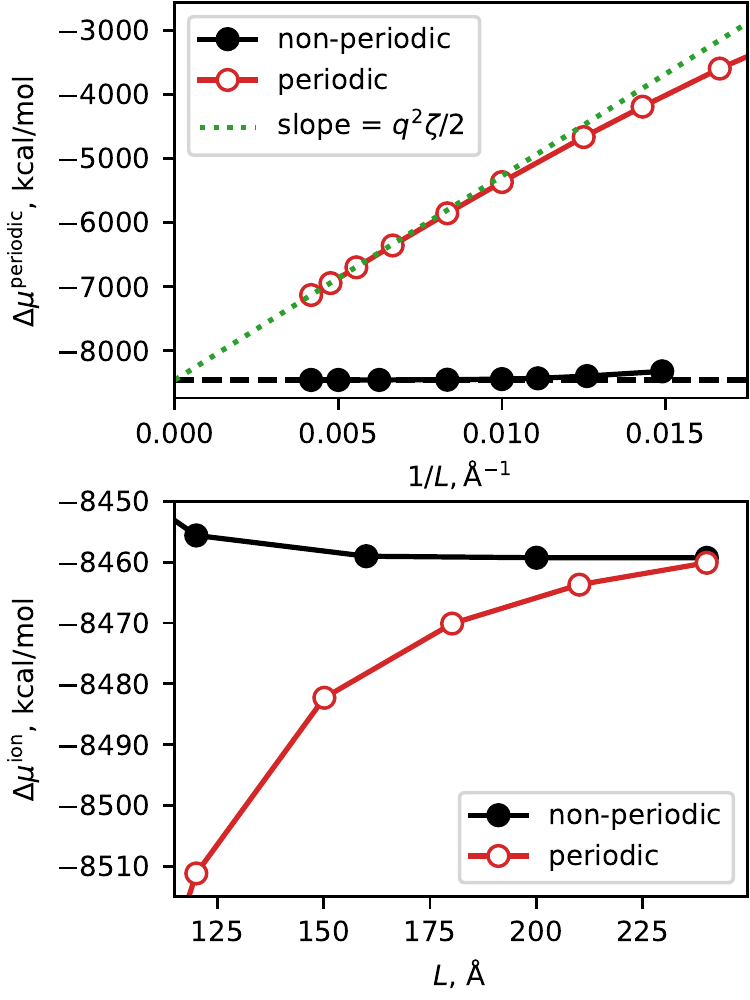}\caption{\label{fig:wigner}Variation of solute excess chemical potential with
respect to cell size. A single sarcin-ricin RNA chain, taken from
PDB ID 480d, is placed in cubic boxes of varying size. The solvent
is 0.1M NaCl in water, with a grid spacing of 0.5 Å. Top: original
results, plotting $\Delta\mu^{\text{periodic}}$; the green line has
a slope of $q^{2}\zeta/2$. Bottom: the periodic result is corrected
to $\Delta\mu^{\text{ion}}$ via Eq. \ref{eq:extrapolate}, and shown
for large box sizes.}
\end{figure}

The thermodynamic quantity of most direct interest is the excess chemical
potential, $\Delta\mu$, since this (when added to the potential energy
of the solute alone) creates the potential of mean force that is used
when applying 3D-RISM as an implicit solvent model. As discussed above,
for a solute with a net charge, the periodic model we use has a uniform
background charge to neutralize the system. A periodic system with
charged molecules and such a uniform background charge is often called
a ``Wigner lattice'', and the effects of periodicity can be computed
and removed, in order to facilitate comparison to comparable non-periodic
calculations. For a cubic cell, the result for a single ion, $\Delta\mu^{\text{ion}}$,
is related to the periodic result as follows:\citep{Lynden-Bell99}
\begin{equation}
\Delta\mu^{\text{ion}}=\Delta\mu^{\text{periodic}}-q^{2}\zeta/2L\label{eq:extrapolate}
\end{equation}
where $q$ is the net charge on the solute, $L$ is the box length,
and $\zeta=2.837$. Fig. \ref{fig:wigner} shows results for a 27-nucleotide
RNA stem-loop with a net solute charge $q$ of -26. The comparison
is to parallel calculations with the existing non-periodic 3D-RISM
codes in Amber. The upper plot illustrates the near-linear dependence
on $1/L$ expected from Eq. \ref{eq:extrapolate}; the lower plot
directly compares $\Delta\mu^{\text{ion}}$ for periodic and non-periodic
codes. In the limit of large box sizes, the two results converge to
the same value (to within 1 kcal/mol at $L=\unit[240]{\mathring{A}}$),
but the non-periodic code is much less sensitive to box size. This
is expected, since the non-periodic result includes an ``asymptotic''
contribution that estimates contributions beyond the box used for
the convolution; this is quite an accurate estimate that provides
reasonably converged results even for modest box sizes. For this reason,
the use of the periodic code for non-periodic problems is not an attractive
option, at least at present. Nevertheless, the existing non-periodic
codes have been well-tested for many types of problems, and the convergence
illustrated in Fig. \ref{fig:wigner} provides evidence for the correctness
of the new periodic implementation.

Another feature of interest, beyond thermodynamics, lies in the solvent
distribution itself. Quantities like the excess number of ions (or
water molecules) around a charged solute can be measured experimentally\citep{Leipply09,Bai07,Gebala_JAmChemSoc_2015_v137_p14705,Pabit10},
and compared with computations. These distributions converge much
more quickly with box size or grid spacing than does $\Delta\mu$
itself. Table \ref{tab:gradient-accuracy-1} gives such values for
the sarcin-ricin RNA in a mixed salt with Mg\textsuperscript{2+},
K\textsuperscript{+} and Cl\textsuperscript{-} ions. Going from
a grid spacing of 0.75 Å to one of 0.25 Å changes $\Delta\mu$ by
39 kcal/mol, whereas the excess number of ions changes hardly at all,
even the excess number of waters changes by only 0.2\%.

\subsection{Solvent distributions in small molecule crystals\label{subsec:xray}}

\begin{figure*}[t]
\begin{centering}
\includegraphics[width=0.7\textwidth]{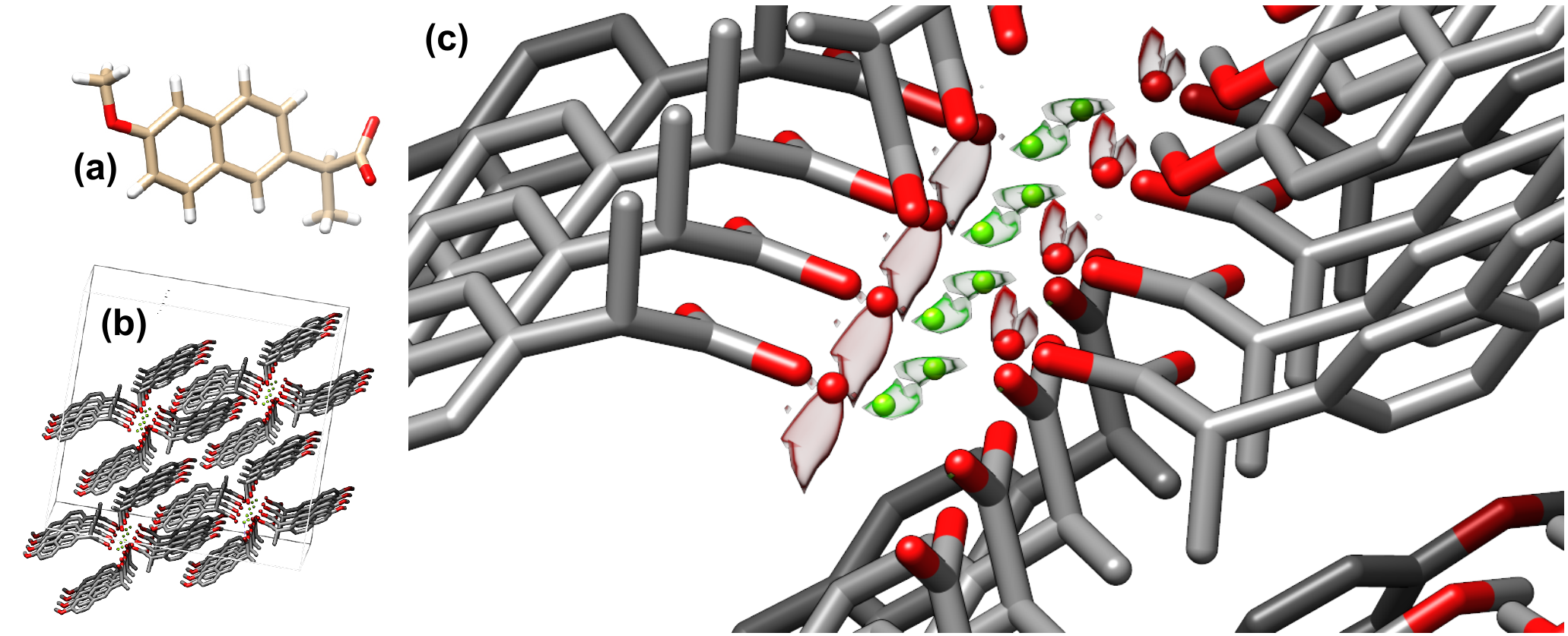}
\par\end{centering}
\caption{(a): Chemical structure and (b): crystal structure (b) of naproxen.H\protect\textsubscript{2}0.Mg\protect\textsuperscript{2+};
(c): solvent density contours for water (red) and Mg\protect\textsuperscript{2+}
(green), with colored spheres showing the locations of localized water
and ions as assigned in the refinement process.\label{fig:naproxen}}
\end{figure*}
One of the key advantages of an atom-based solvent model like 3D-RISM,
compared to continuum implicit solvent models, is that a thermally-averaged
solvent distribution profile (on a 3D-grid) is available for each
solvent component. A simple small-molecule example is the non-steroidal
anti-inflammatory drug naproxen, whose crystal structure (CCDC entry
ANOMEW\citep{Friscic11}) as a hydrate with water and Mg\textsuperscript{2+}
is shown in Figure \ref{fig:naproxen}.  The solvent density contours
from 3D-RISM closely match the electron density distributions from
X-ray crystallography. This may not be surprising this this case,
since the solvent channel is narrow, but offers prospects for analyses
of the many polymorphs of naproxen that have different amounts of
waters and cations, sometimes with clear evidence of disordered solvent.
Similar predictions are available for biomolecules, such as for the
RNA crystals discussed below; but there it is more difficult to evaluate
the accuracy of the 3D-RISM results, since only a small percentage
of the ions and water molecules that must be present in the crystal
can be located in electron density maps.

One way to evaluate the quality of the predicted solvent distributions
is to use them (in combination with atomic models for the solute molecules)
to compute X-ray scattering intensities that can be compared to those
observed from X-ray crystallography. Since atomic models for macromolecules
almost never reproduce experimental X-ray scattering amplitudes to
within experimental data (a feature that is sometimes called the ``R-factor
gap''\citep{Holton14}), we compare results using 3D-RISM to the
standard ``flat'' solvent models employed in conventional crystallographic
refinement.

\begin{figure}[t]
\begin{centering}
\includegraphics{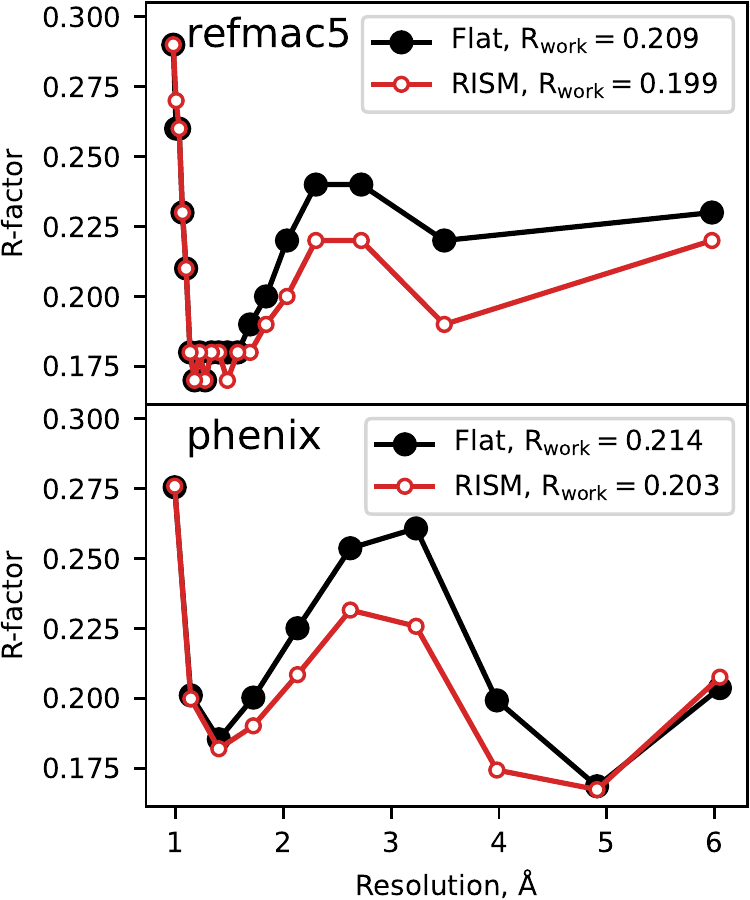}
\par\end{centering}
\caption{Refinement of 1AHO\citep{Smith97a} in (top) \emph{refmac5} and (bottom)
\emph{phenix }using a flat solvent density model and 3D-RISM.\label{fig:wat-1aho}}
\end{figure}

Results are shown in Fig. \ref{fig:wat-1aho} and Tables \ref{tab:rfree2}
to \ref{tab:rfree3}. Refinement calculations were performed using
two popular macromolecular refinement codes, \emph{refmac5}\citep{Murshudov11}
and \emph{phenix}\citep{Liebschner19}. These two codes give broadly
similar results, but differ in details of how the flat solvent model
is implemented and how reflections are binned by resolution and subsequently
scaled. The 3D-RISM solvent density maps were computed using the deposited
solute atomic models (keeping only the most highly occupied alternate
conformations) with solvent molecules removed. During refinement,
the solvent density is held constant (except for overall scaling and
overall B-factors, which are refined), and the atomic positions and
B-factors of the solute are modified to achieve best agreement with
the observed diffraction intensities. We used 40 refinement cycles
for \emph{refmac5} starting from the deposited solute atomic model.
Parallel refinements were carried out using the default, ``flat'',
solvent density model. The \emph{phenix.refine }package does not have
a fully comparable capability, but we can compare 3D-RISM and flat
bulk-solvent models for the deposited solute atomic model.

Fig. \ref{fig:wat-1aho} shows results for a 64-residue scorpion toxin
protein, PDB code 1AHO. There is overall drop of about 1\% between
the flat and 3D-RISM solvent models, with about a 2\% improvement
in resolutions between 2 and 4 Å, whereas there is little difference
at lower and higher resolutions. This is not an insignificant improvement
(given that there are no new adjustable parameters) and provides a
benchmark example for other solvent models, such as those based on
other closures or on MD simulations: better solvent models should
yield lower R-factors. For now, this calculation only provides better
``statistics''; this solvent model would need to be integrated into
a refinement algorithm to see what effect it would have on the final
atomic model. (Such studies will be reported elsewhere.) It is likely
that improved models may involve some combination of explicit water
molecules (placed into locations identified in the electron density
map) and a 3D-RISM model for the remaining (``disordered'' or ``bulk'')
solvent. These more complex models have more adjustable parameters,
which will have to be balanced against improvements in the resulting
R-factors.

\begin{table*}[t]
\begin{ruledtabular}
\begin{centering}
\begin{tabular}{lcccccc}
Protein & scorpion-toxin & GB3 & myoglobin & lysozyme & lysozyme & cyclophilin\tabularnewline
\hline 
PDB ID/resol. & 1AHO/0.96 & 2IGD/1.10 & 1BZR/1.15 & 4LZT/0.95 & 2LZT/1.97 & 4YUL/1.42\tabularnewline
flat (Refmac) & .209/.214 & .220/.233 & .200/.208 & .196/.205 & .167/.216 & .201/.224\tabularnewline
3D-RISM & .199/.211 & .213/.224 & .194/.206 & .190/.197 & .154/.201 & .185/.202\tabularnewline
\end{tabular}
\par\end{centering}
\end{ruledtabular}

\caption{Bulk solvent models with a single protein configuration; each block
shows R/Rfree after 40 cycles of \emph{refmac5} refinement.\label{tab:rfree2}}
\end{table*}

\begin{table}[h]
\begin{ruledtabular}
\begin{centering}
\begin{tabular}{lccc}
RNA & pseudoknot & sarcin-ricin loop & hammerhead\tabularnewline
\hline 
PDB ID/resol. & 2A43/1.34 & 480D/1.50 & 2QUS/2.40\tabularnewline
flat (Refmac) & .223/.261 & .192/.216 & .206/.255\tabularnewline
3D-RISM & .208/.229 & .175/.208 & .186/.234\tabularnewline
\end{tabular}
\par\end{centering}
\end{ruledtabular}

\caption{Bulk solvent models with a single RNA configuration; each block shows
R/Rfree after 40 cycles of \emph{refmac5} refinement.\label{tab:rfree3}}
\end{table}

Tables \ref{tab:rfree2} and \ref{tab:rfree3} show overall drops
in R and Rfree for a selection of small proteins and RNA crystals.
In each case, R and Rfree are improved: on average, the 3D-RISM values
for Rfree are 1.3\% better than when using the default flat solvent
model in \emph{refmac5}. Further studies of alternative bulk solvent
models will be reported elsewhere.

\subsection{Using 3D-RISM as an implicit solvent model for biomolecular crystals\label{subsec:implict-solvent}}

In addition to providing a map of the distribution of solvent molecules
in the crystal lattice, the integral equation approach provides a
solvation free energy and its gradients with respect to solute atomic
positions. This provides an implicit solvent model that can be used
for minimizations or molecular dynamics. This has been found to work
well in non-periodic situations, giving results that are often superior
to numerical Poisson-Boltzmann or generalized Born models.\citep{Onufriev19}
Since there are very few implicit solvent models that work for crowded
periodic systems like molecular crystals, this is an intriguing approach,
in spite of its relatively high computational cost.

\begin{figure}[t]
\begin{centering}
\includegraphics{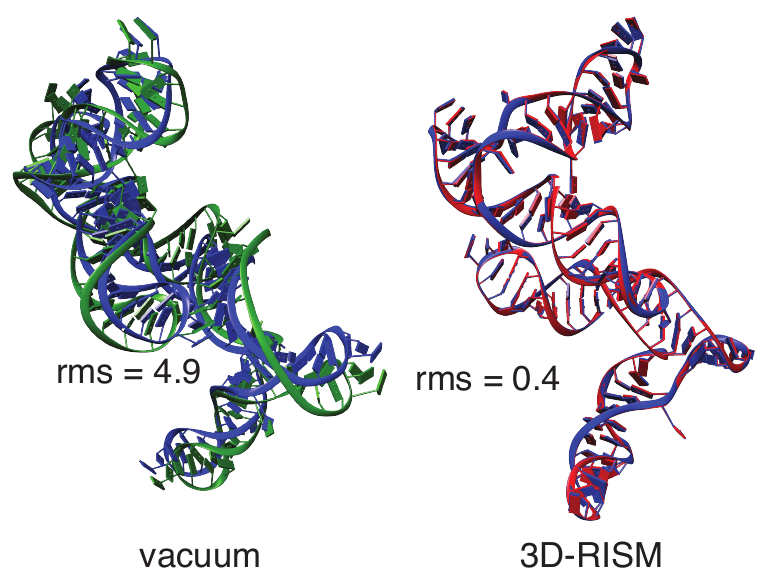}
\par\end{centering}
\caption{\label{fig:2oiu}Blue: experimental structure from X-ray crystallography
(PDB ID 2OIU); red: structure from a 3D-RISM crystal minimization;
green: structure from a crystal minimization with no solvent correction.
RMS gives the root-mean-square deviation (in Å) of all non-hydrogen
atoms from the crystal structure. Only a single chain is shown, but
the calculation included the entire unit cell.}
\end{figure}

\begin{figure}[t]
\begin{centering}
\includegraphics{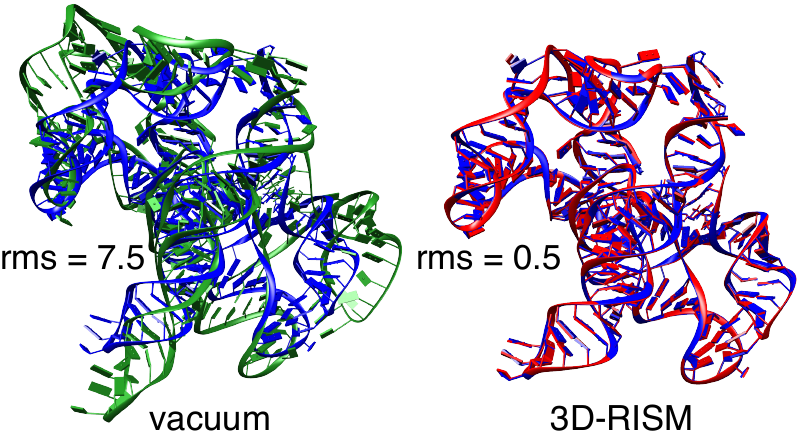}
\par\end{centering}
\caption{\label{fig:1y0q}Same as Fig. \ref{fig:2oiu}, but for PDB code 1Y0Q.}
\end{figure}

The need to include the energetic aspects of solvation is especially
important for nucleic acids crystals, where there are many charged
phosphate groups in close proximity, and generally only a small number
of counter ions are visible in the electron density maps. We consider
two examples here: the L1 ribozyme ligase circular adduct (PDB code
2OIU\citep{robertson2007structural}) and a group I intron product
complex (PDB code 1Y0Q\citep{golden2005crystal}). Figures \ref{fig:2oiu}
and \ref{fig:1y0q} show results of minimization calculations in the
crystal lattice, with and without the 3D-RISM implicit solvent model.
For the smaller 2OIU system (9188 solute atoms), we carried out 1100
steps of conjugate gradient minimization (using the LBFGS algorithm),
followed by 30 steps of truncated-Newton conjugate gradient optimization.
The root-mean-square of the elements of the final gradient was 0.02
kcal/mol-Å, and the energy drop on the final step of truncated-Newton
optimization was 0.3 kcal/mol. For 3D-RISM with a 1.0 Å grid spacing,
each energy evaluation took 13 sec., using 16 MPI threads on a single
Xeon Gold 6230 CPU running at 2.10 GHz. The larger 1Y0Q system (60,288
solute atoms) was minimized for 400 steps of conjugate gradient minimization,
with a final RMS gradient of 0.02 kcal/mol. Here each energy evaluation
required 9 minutes of time on 16 threads on a single CPU.

Figures \ref{fig:2oiu} and \ref{fig:1y0q} show superpositions of
a single RNA chain, even though the simulations themselves included
a full unit cell that is periodically replicated. In both examples,
it is clear that the lack of solvent screening of the phosphate-phosphate
interactions in the ``no solvent model'' minimizations results in
an expansion of the system, even within the constraints of the crystal
lattice, whereas the 3D-RISM calculations show excellent fidelity
to the experimental structural models. (It is not enough to just reduce
the net charge on phosphate groups: for 1Y0Q, a ``vacuum'' minimization
where the net charge on each phosphate is reduced from -1.0 to -0.2,
in rough accord with counterion condenstion models, still results
in an RMS shift of 4.7 Å.) In a refinement calculation without the
implicit solvent model, the force-field energies would be fighting
against the Xray restraints, whereas the results of Figures \ref{fig:2oiu}
and \ref{fig:1y0q} suggest that this would be much less true if 3D-RISM
were employed.

The fairly slow timings for 3D-RISM will limit some potential applications,
but need not impede useful results. For example, a typical 10-cycle
refinement run in the \emph{phenix} package of programs\citep{Liebschner19}
typically makes fewer than 300 energy evaluations during the coordinate
refinement steps, so that even a system as large as 1y0q would need
less than 2 days of time, which is not inappropriate for a final refinement
step. (We have begun coding a GPU-enabled version of these models,
and hope that this will provide a significant speed improvement over
the CPU results reported here.)

\begin{table*}
\begin{ruledtabular}
\begin{centering}
\begin{tabular}{ldddd}
 & \multicolumn{1}{c}{1Y0Q} & \multicolumn{1}{c}{phenix\_cdl} & \multicolumn{1}{c}{phenix-amber} & \multicolumn{1}{c}{3D-RISM}\tabularnewline
\hline 
clashscore & 53.7 & 35.4 & 3.7 & 0.9\tabularnewline
RMS(bonds) & 0.008 & 0.011 & 0.017 & 0.015\tabularnewline
RMS(angles) & 1.35 & 2.10 & 3.00 & 2.00\tabularnewline
molprobity score & 3.35 & 3.18 & 2.31 & 1.91\tabularnewline
pucker outliers (\%) & 8.6 & 8.6 & 10.7 & 8.2\tabularnewline
angle outliers (\%) & 0.7 & 0.7 & 9.4 & 2.0\tabularnewline
average suiteness & 0.492 & 0.414 & 0.307 & 0.574\tabularnewline
R-work & 0.277 & 0.221 & 0.264 & 0.251\tabularnewline
R-free & 0.310 & 0.278 & 0.307 & 0.293\tabularnewline
RMS from deposited & 0.00 & 0.36 & 0.71 & 0.37\tabularnewline
\end{tabular}
\par\end{centering}
\begin{centering}
\par\end{centering}
\end{ruledtabular}

\caption{\label{tab:rismrefine}Results for several test refinements of 1Y0Q.
The first seven rows come from the molprobity program;\citep{Chen10g}
the root-mean-square (RMS) change from the deposited structure is
computed for all non-hydrogen atoms.}
\end{table*}

As an example, we show in Table \ref{tab:rismrefine} results for
several crystallographic refinement calculations for the group I intron,
PDB code 1Y0Q. The diffraction data here are only at 3.6 Å resolution,
so many structural details are not well-determined by the X-ray data
alone. The first column shows the deposited results and gives statistics
from the \emph{molprobity} program.\citep{Chen10g} The next two columns
show parallel refinements (starting from the deposited structure)
using phenix: the ``phenix\_cdl'' column uses the default geometric
restraints from its Conformational Dependent Library, which are largely
similar to conventional Engh-Huber restraints. The ``phenix\_amber''
column replaces the cdl restraints with forces from the Amber force
field, as described elsewhere.\citep{Moriarty20} This force field
model has no implicit solvent contribution, and hence no charge-screening
effects. The final column adds in the 3D-RISM model as in Figure \ref{fig:1y0q};
we used in-house codes to carry out the coordinate refinements, and
\emph{phenix.refine }for isotropic B-factor refinements, alternating
cycles of 150 refinement steps of coordinate refinement with 5 macro-cycles
of B-factor optimization.

The overall results are in general agreement with earlier studies
on proteins.\citep{Moriarty20} The use of a force field greatly reduces
the number of bad contacts, as evidenced by the clashscore and improves
the overall molprobity score. But the RNA-specific scores for sugar
pucker, sugar angles and ``suiteness'' (a measure of how well sugar-phosphate
torsion angles agree with databases of well-refined structures) get
worse in the phenix-amber results. This is presumably because the
force field itself prefers an expanded structure (Figure \ref{fig:1y0q})
and its gradients are competing with those from the observed structure
factors. The addition of the 3D-RISM model improves all of the structural
features, and reduces the shift away from the deposited structure.
Comparable results for six additional RNA crystals are presented elsewhere.\citep{Gray21a}

It is clear that many more studies will be needed to establish the
generality of these results: in proteins, where charge screening effects
are less important, more than 13,000 such parallel refinements were
carried out to help establish expected behavior.\citep{Moriarty20}
Systems with higher-resolution diffraction data should depend less
on the nature of the geometric restraints than do lower-resolution
structures. But these initial results illustrate what is now possible
in this regard.

\section{Conclusions}

Water molecules and ions around biomolecules often play a crucial
role in function. Analysis of the solvent distributions in biomolecular
crystals can provide an important check on the accuracy of computational
models. Here we present an implementation of the 3D-RISM solvent model
that can be applied to any periodic system, included ``crowded''
systems like crystals, where the majority of space is taken up by
the solute.

In many ways, the periodic version is not a major departure from existing,
non-periodic 3D-RISM codes, since fast Fourier transforms (with a
periodic cell) have always been used to compute the convolutions needed
for the Ornstein-Zernike equation. The machinery to compute the periodic
potential energy was adapted from existing particle-mesh-Ewald (PME)
procedures in molecular dynamics code. But a key advance was required
for charged solutes: a modification of the total correlation function
$h$ is needed (Eq. \ref{eq:hbk}) to account for the implicit neutralizing
potential arising from the PME procedure, and this in turn implies
an extra contribution to the excess chemical potential (Eq. \ref{eq:exchem-hnc-shifted})
that had not been recognized before. This contribution is negligible
for non-periodic systems, but can become important for crowded crystalline
environments. With this correction, analytical expressions for forces
on the solute atoms closely match gradients computed by finite difference,
and the periodic expressions smoothly merge to existing non-periodic
results for a single solute as the size of the periodic cell increases.
Our approach for charged solutes does involve a uniform background
charge distribution (so that $\mu^{\text{PME}}$ can be used in place
of $u$.) This method of unit-cell neutralization is neither physical
nor unique, but does lead to an internally consistent approach with
accurate gradients (Table \ref{tab:gradient-accuracy-1}) and preliminary
results that are promising even for highly-charged systems (Figs.
\ref{fig:1y0q} and \ref{fig:2oiu} and Table \ref{tab:rismrefine}.)

It is clear that much effort will be required to understand the expected
accuracy of this approach and that improvements in potentials and
in closure relations should be examined. The predicted solvent distributions
can be compared to experiment in a variety of ways: by looking at
the locations of ordered waters and ions that can be identified in
density maps derived from Xray crystallography; by comparing computed
and observed Bragg intensities; and (potentially) by comparing predicted
and measured crystal densities (which reflect the total number of
water and ions per unit cell). Use of 3D-RISM as a periodic implicit
solvent model can be tested by molecular dynamics or minimization
calculations in cases where experimental structures are available.
We have provided a few examples of such comparisons here, but many
more are needed. Improvements in efficiency will help to make this
a practical method; porting the codes to a GPU environment is underway.

The periodic 3D-RISM implementation used here will be included in
AmberTools, an open source collection of molecular simulation software,
and may downloaded at \href{https://ambermd.org}{https://ambermd.org}.
The implementation was based upon an existing non-periodic RISM code
that was primarily developed by Tyler Luchko, David Case, and Andriy
Kovalenko \citep{Luchko10}. Extensions to periodic systems were implemented
by Jesse Johnson and George Giambasu, and a more complete description
of the codes is given elsewhere.\citep{Johnson16a}
\begin{acknowledgments}
This work was supported by National Institutes of Health under award
GM122086 and by the National Science Foundation under grants CHE-1566638
and CHE-2018427. We thank Timothy J. Giese for help in the particle-mesh
Ewald procedures used to compute the electrostatic potential and to
extract the resulting forces on atoms, Jesse Johnson for much work
on the initial version of the periodic code.\citep{Johnson16a}, and
Pavel Afonine and James Holton for help with the \emph{phenix} and
\emph{refmac5} bulk solvent analyses.
\end{acknowledgments}

\section*{Data availability}

All data that support the findings of this study are available from
the corresponding author upon reasonable request and can be can be
reproduced with the AmberTools 21 software suite \citep{case2021amber2021}.

\appendix

\section{The excess chemical potential for the PSE-n closure family}

For the partial series expansion of order-$n$ (PSE-$n$) family of
closures \citep{Kast_JChemPhys_2008_v129_p236101}, which includes
the Kovalenko-Hirata (KH) closure,\citep{Kovalenko00a} we have
\[
g_{\gamma}\left(\mathbf{r}\right)=\begin{cases}
\sum_{0}^{n}\frac{\left(t_{\gamma}^{*}\left(\mathbf{r}\right)\right)^{i}}{i!} & t^{*}\left(\mathbf{r}\right)>0\\
\exp\left(t^{*}\right) & t^{*}\left(\mathbf{r}\right)\le0
\end{cases},
\]
which has the bridge function
\[
B_{\gamma}\left(\mathbf{r}\right)=\begin{cases}
-t_{\gamma}^{*}\left(\mathbf{r}\right)+\ln\left(\sum_{0}^{n}\frac{\left(t_{\gamma}^{*}\left(\mathbf{r}\right)\right)^{i}}{i!}\right) & t^{*}\left(\mathbf{r}\right)>0\\
0 & t^{*}\left(\mathbf{r}\right)\le0
\end{cases}
\]
where $t_{\gamma}^{*}\left(\mathbf{r}\right)=-\beta u_{\gamma}\left(\mathbf{r}\right)+h_{\gamma}\left(\mathbf{r}\right)-c_{\gamma}\left(\mathbf{r}\right)$.
Because we have a non-zero bridge function, we must consider the variation
in the general form of the closure, Eq. \ref{eq:ClosureEquation},
\begin{multline*}
\delta h_{\gamma}\left(\mathbf{r}\right)=-g_{\gamma}\left(\mathbf{r}\right)\beta\delta u_{\gamma}^{\text{PME}}\left(\mathbf{r}\right)+g_{\gamma}\left(\mathbf{r}\right)\delta h_{\gamma}\left(\mathbf{r}\right)\\
-g_{\gamma}\left(\mathbf{r}\right)\delta\tilde{c}_{\gamma}\left(\mathbf{r}\right)+g_{\gamma}\left(\mathbf{r}\right)\delta B_{\gamma}\left(\mathbf{r}\right).
\end{multline*}
All but the last was treated in section \ref{subsec:Computing-the-excess}.
For the last term, we have an exact differential,
\begin{multline*}
g_{\gamma}\left(\mathbf{r}\right)\delta B_{\gamma}\left(\mathbf{r}\right)\\
\begin{aligned} & =\begin{cases}
\begin{aligned} & \sum_{0}^{n}\frac{\left(t_{\gamma}^{*}\left(\mathbf{r}\right)\right)^{i}}{i!}\Biggl[-\delta t_{\gamma}^{*}\left(\mathbf{r}\right)\\
 & +\frac{1}{\sum_{0}^{n}\frac{\left(t_{\gamma}^{*}\left(\mathbf{r}\right)\right)^{i}}{i!}}\sum_{0}^{n-1}\frac{\left(t^{*}\left(\mathbf{r}\right)\right)^{i}}{\left(i\right)!}\delta t_{\gamma}^{*}\left(\mathbf{r}\right)\Biggr]
\end{aligned}
 & t^{*}\left(\mathbf{r}\right)>0\\
0 & t^{*}\left(\mathbf{r}\right)\le0
\end{cases}\\
 & =-\delta\frac{\left(t_{\gamma}^{*}\left(\mathbf{r}\right)\right)^{n+1}}{\left(n+1\right)!}\Theta\left(t_{\gamma}^{*}\left(\mathbf{r}\right)\right),
\end{aligned}
\end{multline*}
where $\Theta\left(\right)$ is the Heaviside function and we have
used $\delta\left(t_{\gamma}^{*}\left(\mathbf{r}\right)\right)^{n}=nt_{\gamma}^{*}\left(\mathbf{r}\right)^{n-1}\delta t_{\gamma}^{*}\left(\mathbf{r}\right)$.
Using this result with Eqs. \ref{eq:kirkwood-charging-1}, \ref{eq:kirkwood-charging-integrand-1-1}
and \ref{eq:kirkwood-hc-exact-1} we have
\begin{multline}
\Delta\mu^{\text{PSE-\ensuremath{n}}}=kT\sum_{\gamma}\rho_{\gamma}\int_{V_{\text{cell}}}\frac{\left(h_{\gamma}\left(\mathbf{r}\right)\right)^{2}}{2}-\left(1-\frac{h_{\gamma}^{\text{bk}}}{2}\right)\tilde{c}_{\gamma}\\
-\frac{h_{\gamma}\left(\mathbf{r}\right)\tilde{c}_{\gamma}\left(\mathbf{r}\right)}{2}-\frac{\left(t_{\gamma}^{*}\left(\mathbf{r}\right)\right)^{n+1}}{\left(n+1\right)!}\Theta\left(t_{\gamma}^{*}\left(\mathbf{r}\right)\right)\,d\mathbf{r}.\label{eq:exchem-psen-shifted}
\end{multline}
As with the HNC closure, this expression is the same as the usual
expression \citep{Kast_JChemPhys_2008_v129_p236101} except for an
additional term of $-kT\sum_{\gamma}\rho_{\gamma}\int_{V_{\text{cell}}}-\frac{1}{2}h_{\gamma}^{\text{bk}}\tilde{c}_{\gamma}\,d\mathbf{r}$.

\bibliography{periodic-rism2}
\end{document}